\documentclass[a4paper,11pt]{article}
\pdfoutput=1 

\usepackage{jinstpub} 

\title{\boldmath Performance of the prototype beam drift chamber for LAMPS at RAON with proton and $^{12}$C beams}

\author[a]{H. Kim,}
\author[b]{Y. Bae,}
\author[a]{C. Heo,}
\author[a]{J. Seo,}
\author[c]{J. Hwang,}
\author[a,1]{D. H. Moon\note{Corresponding author.}}
\author[d]{D. S. Ahn,}
\author[c,e]{J. K. Ahn,}
\author[b]{J. Bae,}
\author[f]{J. Bok,}
\author[g]{Y. Cheon,}
\author[c,e]{S. W. Choi,}
\author[c,e]{S. Do,}
\author[c,e]{B. Hong,}
\author[h]{S.-W. Hong,}
\author[d]{J. Huh,}
\author[i]{S. Hwang,}%
\author[c,e]{Y. Jang,}
\author[c,e]{B. Kang,}
\author[c,e]{A. Kim,}
\author[b]{B. Kim,}
\author[f]{C. Kim,}
\author[j]{E.-J. Kim,}
\author[g]{G. Kim,}
\author[k]{G. Kim,}
\author[c,e]{J. Kim,}
\author[e,k]{J. Kim,}
\author[l]{S. H. Kim,}
\author[f]{Y. Kim,}
\author[c,e]{Y. J. Kim,}
\author[g]{Y. Kim,}
\author[h]{Y. J. Kim,}
\author[k]{M. Kweon,}
\author[h]{C. Lee,}
\author[c,e]{H. Lee,}
\author[b]{H. Lee,}
\author[h]{H. Lee,}
\author[c,e]{J. Lee,}
\author[c,e]{J. Lee,}
\author[c,e]{J.-W. Lee,}
\author[d]{J. W. Lee,}
\author[k]{S. H. Lee,}
\author[g]{S. Lee,}
\author[c,e]{S. Lee,}
\author[f]{S. Lim,}
\author[c,e]{S. H. Nam,}
\author[c,e]{J. Park,}
\author[h]{and T. Shin}

\affiliation[a]{Department of Physics, Chonnam National University, Gwangju, Republic of Korea, 61186}
\affiliation[b]{Department of Physics, Sungkyunkwan University, Suwon, Republic of Korea, 16419}
\affiliation[c]{Department of Physics, Korea University, Seoul, Republic of Korea, 02841}
\affiliation[d]{Center for Exotic Nuclear Studies (CENS), Institute for Basic Science, Daejeon, Republic of Korea, 34126}
\affiliation[e]{Center for Extreme Nuclear Matters (CENuM), Korea University, Seoul, Republic of Korea, 02841}
\affiliation[f]{Department of Physics, Pusan National University, Busan, Republic of Korea, 46241}
\affiliation[g]{Department of Physics and Astronomy, Sejong University, Seoul, Republic of Korea, 05006}
\affiliation[h]{Institute for Rare Isotope Science (IRIS), Institute for Basic Science, Daejeon, Republic of Korea, 34300}
\affiliation[i]{Korea Research Institute of Standards and Science (KRISS), Daejeon, Republic of Korea, 34113}
\affiliation[j]{Division of Science Education, Jeonbuk National University, Jeonju, Republic of Korea, 54896}
\affiliation[k]{Department of Physics, Inha University, Incheon, Republic of Korea, 22212}
\affiliation[l]{Department of Physics, Kyungpook National University, Daegu, Republic of Korea, 41566}

\emailAdd{dhmoon@chonnam.ac.kr}

\abstract{
Beam Drift Chamber (BDC) is designed to reconstruct the trajectories of incident rare isotope beams provided by RAON (Rare isotope Accelerator complex for ON-line experiments) into the experimental target of LAMPS (Large Acceptance Multi-Purpose Spectrometer). 
To conduct the performance test of the BDC, the prototype BDC (pBDC) is manufactured and evaluated with the high energy ion beams from HIMAC (Heavy Ion Medical Accelerator in Chiba) facility in Japan. Two kinds of ion beams, 100 MeV proton, and 200 MeV/u $^{12}$C, have been utilized for this evaluation, and the track reconstruction efficiency and position resolution have been measured as the function of applied high voltage.
This paper introduces the construction details and presents the track reconstruction efficiency and position resolution of pBDC. 
}


\begin{document}
\maketitle
\flushbottom

\section{Introduction}
\label{sec:intro}

A facility dedicated to generate a radioactive ion beam (RIB) holds the potential to offer crucial insights into the various aspects of nuclear matter and exotic nuclei characterized by exceptional neutron-to-proton ratios ($N/Z$). The RIB facility has found extensive use owing to its capacity to investigate fundamental questions, including the origins of matter, nucleus structure, and stability, as well as astrophysical phenomena such as supernovae and neutron stars. In general, the binding energy of a nucleus per nucleon is described using the empirical Bethe-Weisz\"{a}cker formula~\cite{ref1}:

\begin{equation}
\frac{E}{A} = E(\rho, \delta = 0) + E_{\rm{sym}}(\rho)\delta^2 + \mathcal{O}(\delta^4) + \dots
\label{eqn:nbind}
\end{equation}
where $\rho = \rho_{n} + \rho_{p}$ is the total baryon density and $\delta = (\rho_{n} - \rho_{p})/\rho$ is the isospin-asymmetric-fraction parameter with $\rho_{n}$ and $\rho_{p}$ being the neutron and proton densities, respectively. 
In eq.~\ref{eqn:nbind}, $E_{\rm{sym}}(\rho)$, the coefficient of $\delta^{2}$, is called the density-dependent nuclear symmetry energy, which is the difference between the energy of pure neutron matter and that of nuclear matter possessing symmetry in isospin degrees of freedom ($N = Z$).

The symmetry energy ($E_{\rm{sym}}$) in this equation of state (EoS) for nuclear matter plays a pivotal role in advancing our comprehension of the many-body theory governing strongly interacting systems~\cite{ref2, ref3}. The primary aim of the LAMPS experiment is to advance our understanding of nuclear symmetry energy by revealing previously unknown knowledge. 

The LAMPS system~\cite{ref4,ref5,ref6,ref7} is installed in the high energy experimental hall at the Rare isotope Accelerator complex for ON-line experiments (RAON), which is the new RIB accelerator being presently under construction in Korea~\cite{ref8,ref9} combining the ISOL (Isotope Separation On-Line) and the IF (In-flight Fragmentation). 
It possesses the capability to deliver neutron-rich rare isotopes such as $^{132}\rm{Sn}$ beams up to 250 MeV/u of intensity as large as $10^8$ pps. Utilizing such a beam, the expected baryon density could reach up to about factor two of normal nuclear saturation density, $\rho_0$ ($\sim$0.16 $\rm{fm}^{-3}$)~\cite{ref7}. 
Then, it will be possible to shed light to determine the trend of the symmetry energy beyond normal nuclear saturation density.

\begin{figure}[htbp]
\centering
\includegraphics[width=.8\textwidth]{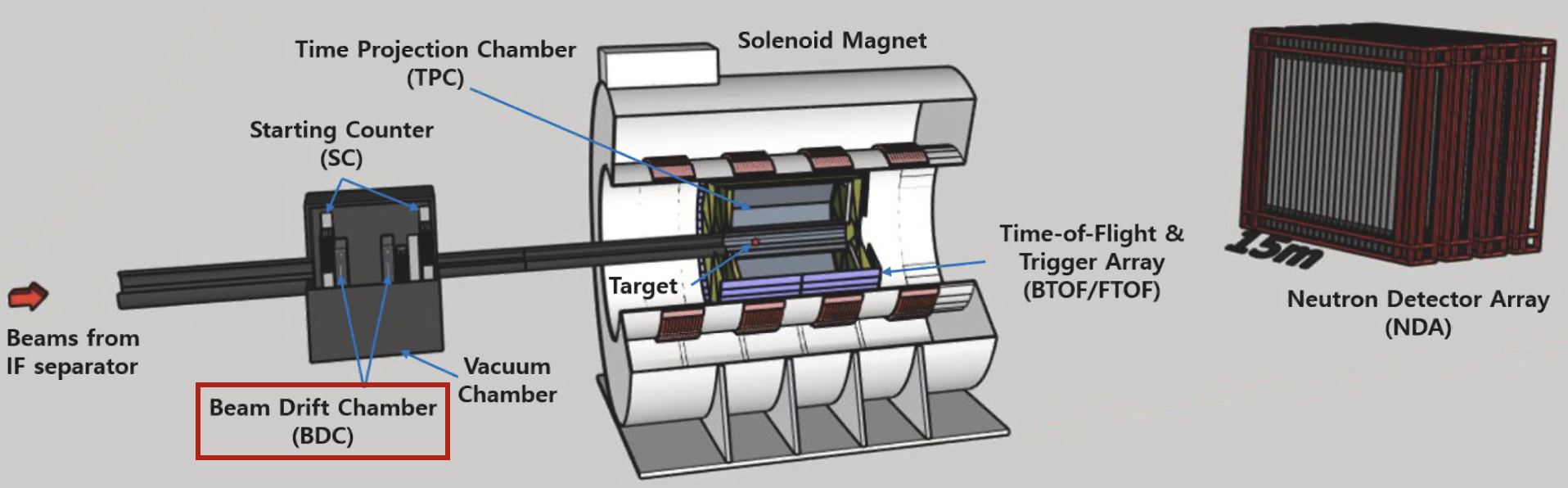}
\caption{\label{fig:fig1} LAMPS outline at RAON. The beams from the IF separator are entering from the left. The location of the BDC is highlighted~\cite{ref7}.}
\end{figure}

The LAMPS system was developed to simultaneously detect charged particles and neutrons, employing event-by-event measurements of collision centrality and reaction plane. The schematic layout of the LAMPS system is illustrated in figure~\ref{fig:fig1}, with beams entering from the left through the IF separator. The foremost component of LAMPS is the beam diagnostic vacuum chamber, housing the Starting Counters (SC) and the Beam Drift Chambers (BDC). Adjacent to the vacuum chamber is the superconducting solenoid magnet, hosting the Time-Projection Chamber (TPC), as well as the Barrel and Forward Time-of-Flight systems (BToF and FToF, respectively). Finally, the Neutron Detector Array (NDA) is positioned in the forward region~\cite{ref4,ref5,ref10}.

Of all the elements within LAMPS, the BDC housed in the vacuum chamber was designed to trace the trajectories of incoming RI beams from their entry point in the beam line to their target within the charged particle detection system.
The design goal of the integrated BDC is detection efficiency of more than 99\% and position resolution of less than 100 $\rm{{\mu m}}$.

This paper delves into a detailed exploration of the design and assembly processes of the prototype BDC (pBDC), crafted specifically for the performance evaluation of the authentic BDC. The performance study results will be presented by executing 100 MeV proton and 200 MeV/u $^{12}$C ion beams at the Heavy Ion Medical Accelerator in Chiba (HIMAC) facility at QST (National institutes for Quantum Science and Technology) located in Chiba, Japan.

\section{Experimental setup}
\subsection{Production of the prototype BDC}
\label{sec:const}

\begin{figure}[htbp]
\centering
\includegraphics[width=.7\textwidth]{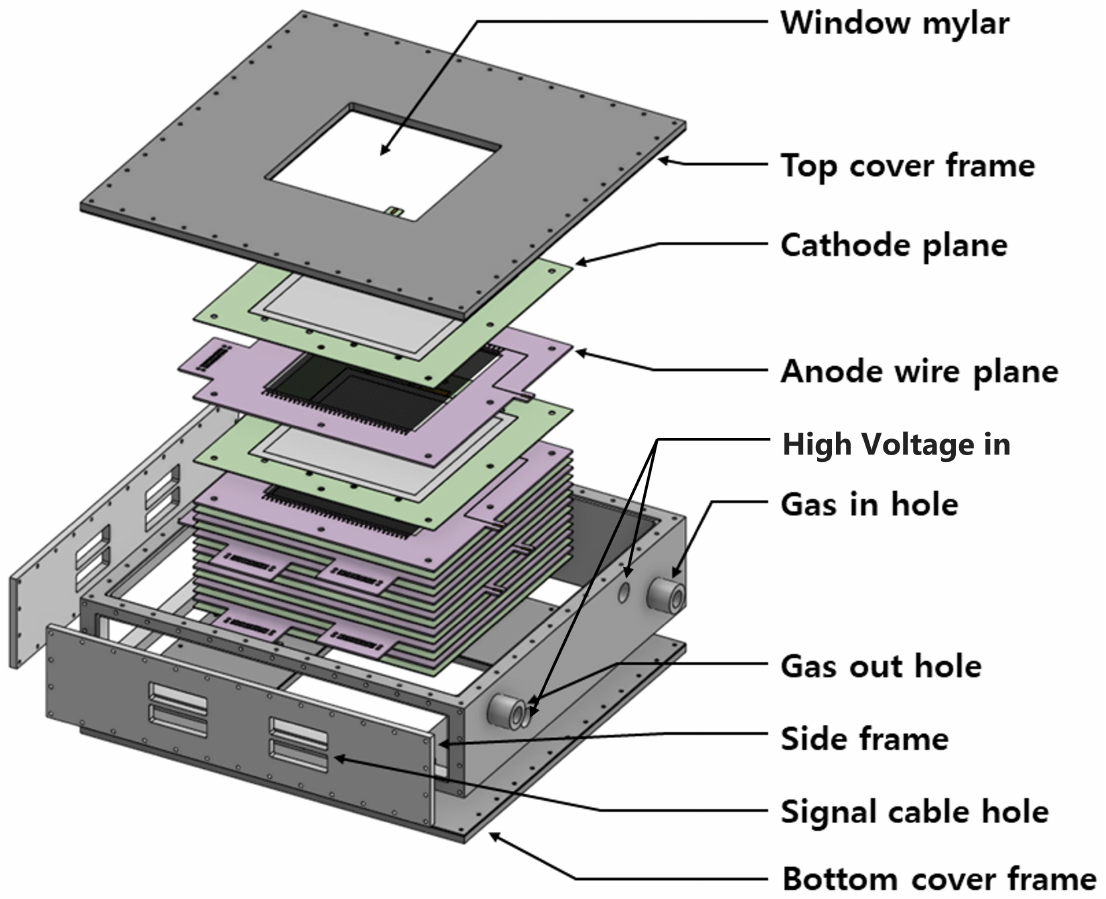}
\caption{\label{fig:fig2} Assembly view of the prototype BDC.}
\end{figure}

The pBDC is the Walenta-type gaseous drift chamber~\cite{ref11} optimized for minimizing beam energy loss and providing good position resolution of beam trajectories.
The detailed assembly process is shown in the schematic diagram of figure~\ref{fig:fig2}.
The chamber box frame is constructed using stainless steel and contains nine cathode planes made of aluminized material along with eight signal anode planes.
From the bottom, the cathode and anode planes are crossed and stacked with a space of 2.5 mm.
The square-shape windows are placed at both the top and bottom side of the frame, covered with 50 $\rm{{\mu m}}$ thick mylar for provided beams passing through.
Considering the expected beam size, the sensitive active area is designed with the dimension of $\rm{160~\times~160~mm^{2}}$. 

\begin{figure}[htbp]
\centering
\includegraphics[width=.4\textwidth]{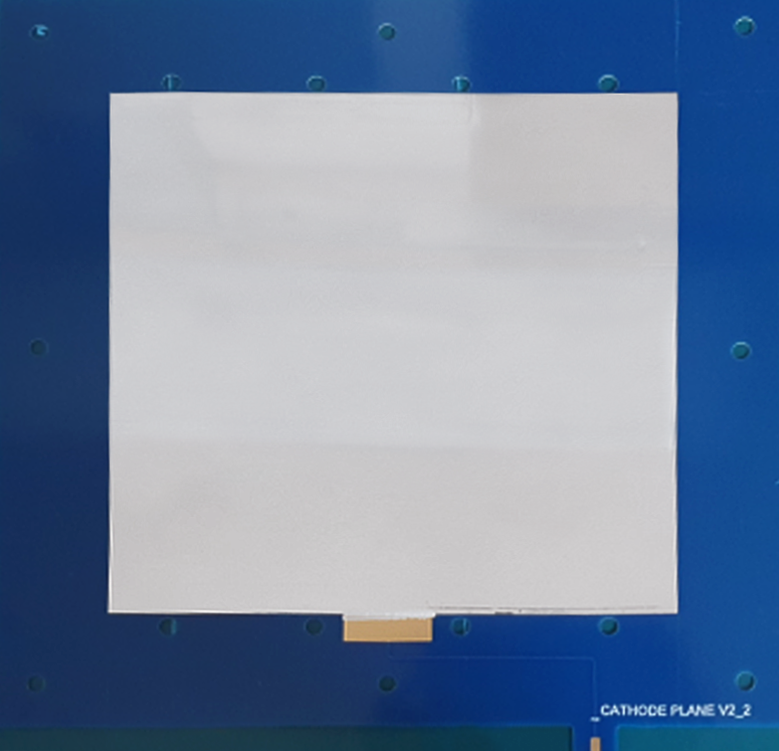}
\includegraphics[width=.4\textwidth]{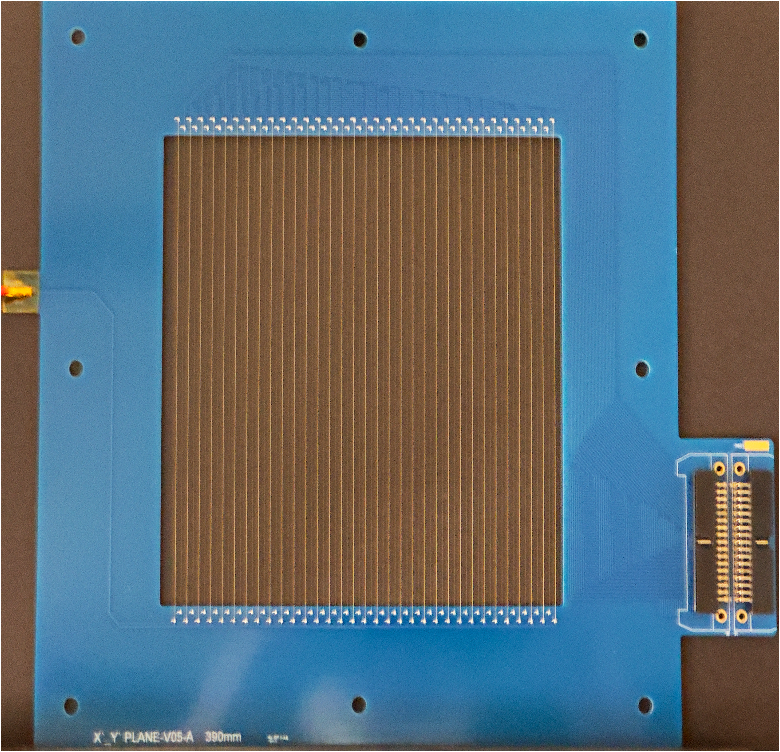}
\caption{\label{fig:fig3} Cathode (left) and anode (right) planes of the pBDC.}
\end{figure}

Figure~\ref{fig:fig3} illustrates the anode and cathode planes of the pBDC.
Each anode plane is made of 33 potential wires and 32 signal wires soldered between the upsides and downsides of the Printed Circuit Board (PCB).
The wires, made of gold-plated tungsten, have diameters of 80 and 20 $\rm{\mu m}$ for potential and signal wires, respectively.
Wire tension is controlled during the soldering process using weights of 100 g for the potential wire and 50 g for the signal wire.

\begin{table}[htb!]
\begin{center}
    \centering
    \caption{Specification of prototype BDC.}
    \begin{tabular}{l|l}
\hline
Parameter & Value \\
\hline\hline
Signal wire & 20 $\rm{{\mu m}}$, Au--W \\
Potential wire & 80 $\rm{{\mu m}}$, Au--W \\
Anode configuration & $XX'XX'YY'YY'$ (total 8 layers) \\
Cathode & 6 $\rm{\mu m}$, Aluminized mylar, 9 layers \\
Cell size & 5 mm (drift length 2.5 mm) \\
Active area & $\rm{160~\times~160~mm^{2}}$ \\
Square-shape windows & 50 $\rm{{\mu m}}$, mylar \\
Number of channels & 256 (32 wires/plane $\times$ 8 planes) \\
Operation gas & P-10 (Ar (90$\%$) + $\rm{CH_{4} (10\%)}$) at 1 atm \\
\hline
    \end{tabular}
    \label{tab:tab1}
\end{center}
\end{table}

\begin{figure}[htb!]
\centering
\includegraphics[width=.8\textwidth]{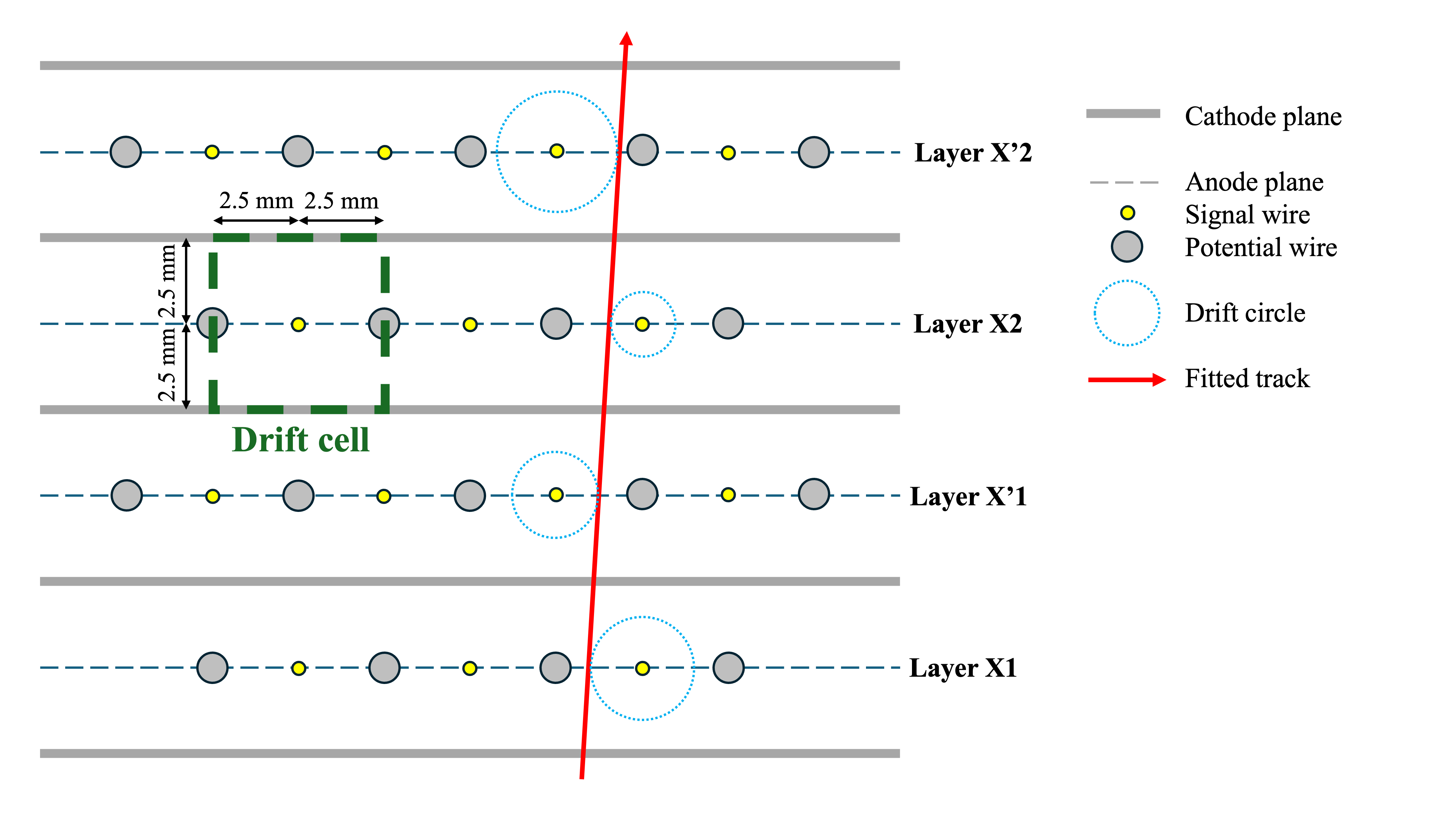}
\caption{\label{fig:fig4} Diagrammatic view of the cathode and anode layers in pBDC.}
\end{figure}

\begin{figure}[htb!]
\centering
\includegraphics[width=.55\textwidth]{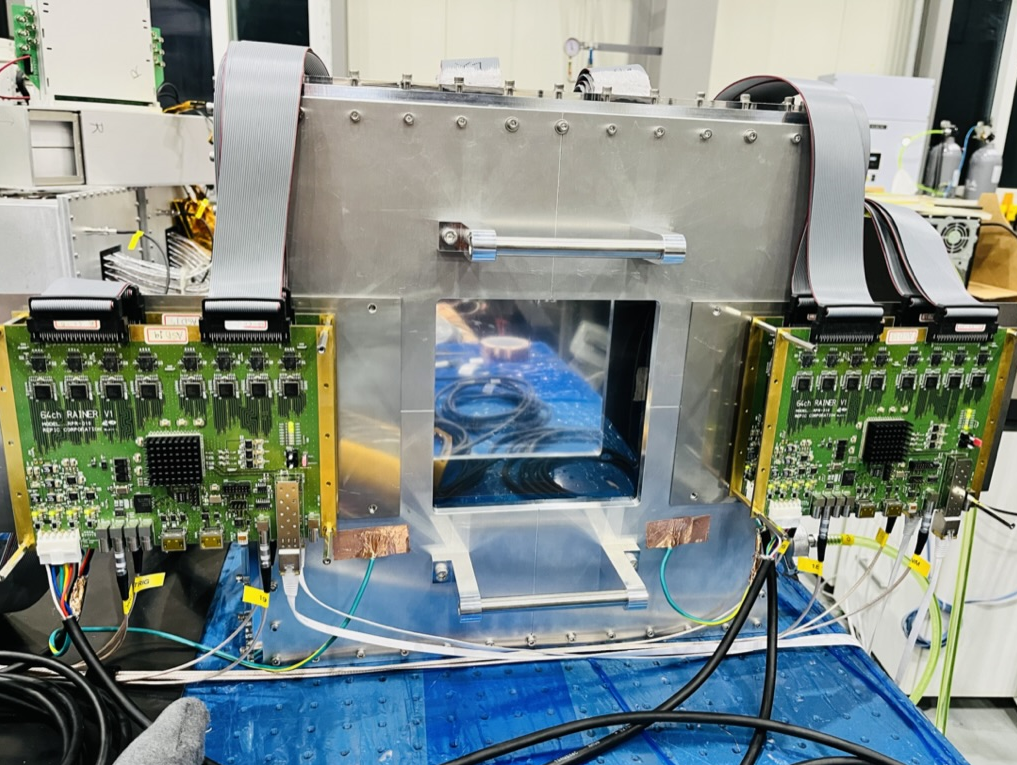}
\caption{\label{fig:fig5} Assembled pBDC with ASD boards.}
\end{figure}

The drift cell width is set to 5.0 mm, determined by the distance of 2.5 mm between the signal and the adjacent potential wires and also between the wire and cathode planes, as displayed in figure~\ref{fig:fig4}.
The signal plane follows the scheme $XX'XX'YY'YY'$ from the upstream, where the $X'$ ($Y'$) plane wires are shifted by 2.5 mm from their counterparts in the $X$ ($Y$) plane, enhancing position resolution.
Soldering points on the anode plane connect to the Insulation Displacement Connector (IDC) via internal lines, and flat cables link the IDC to the Amplifier-Shaper-Discriminator (ASD) board through the protection board to protect the ASD board from unexpected high currents from the wires.
Within each cathode layer, a double-sided aluminized mylar with a thickness of 6 $\rm{\mu m}$ covers the window centered in the PCB frame.
Figure~\ref{fig:fig5} shows the assembled pBDC with ASD boards.
The specifications of the pBDC are summarized in table~\ref{tab:tab1}.

\subsection{Beam test setup}
\label{sec:setup}

The performance of pBDC is evaluated using the beam provided by the HIMAC 
which is an accelerator primarily designed for cancer therapy and capable of accelerating various ions, including proton and $^{12}$C ions, to energies ranging from 100 to 800 MeV/u, depending on the species of ions.
100 MeV proton and 200 MeV/u $^{12}\rm{C}^{+6}$ ion beams were utilized for our test.

\begin{figure}[htb!]
\centering
\includegraphics[width=1.\textwidth]{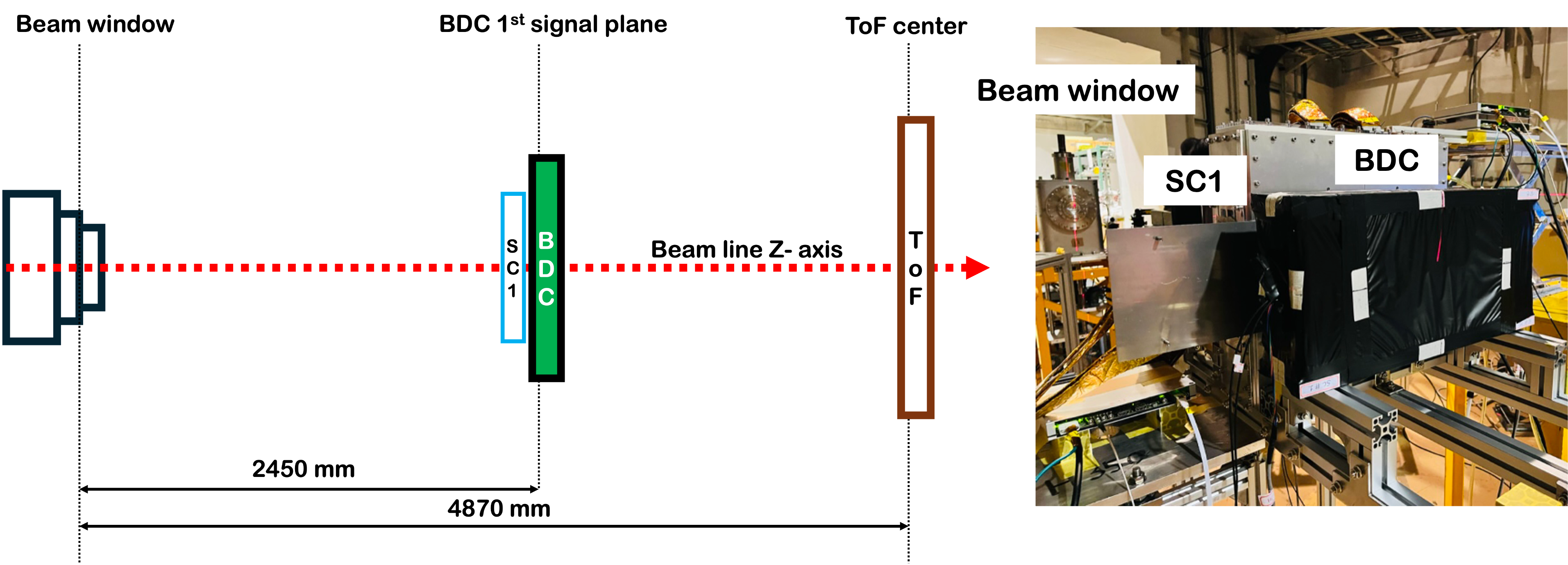}
\includegraphics[width=1.\textwidth]{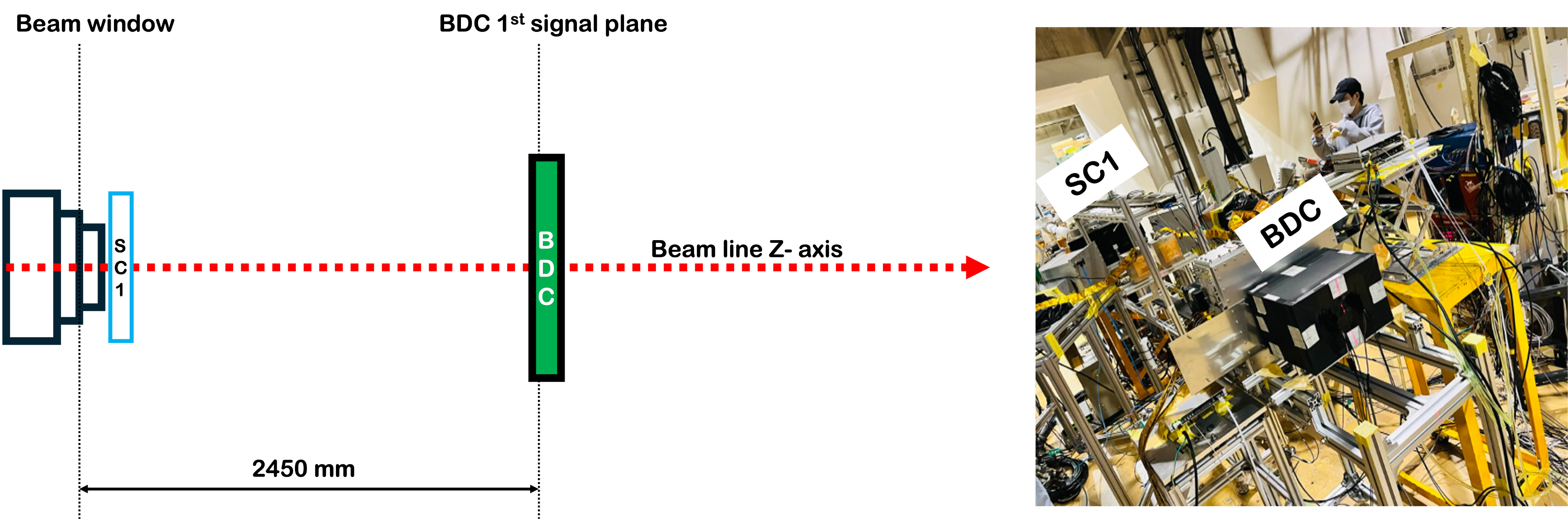}
\caption{\label{fig:fig6} Experimental setup schematic view (left) and snapshot (right) for proton (up) and $^{12}$C beam (down).}
\end{figure}

The experimental configuration with snapshots corresponding to proton and $^{12}$C ion beam setups is demonstrated in figure~\ref{fig:fig6}.
For proton case, the coincidence signal between forward SC (SC1) and downstream ToF was employed as the trigger signal.
In the cases of the $^{12}$C beam, SC1 was located nearest to the beam window, followed by the ToF detector aligned with the beam axis.
In any case, the time information from only SC1 was used as the triggered time, considering coordination with other experimental tests.

\begin{figure}[htb!]
    \centering
    \includegraphics[width=.8\textwidth]{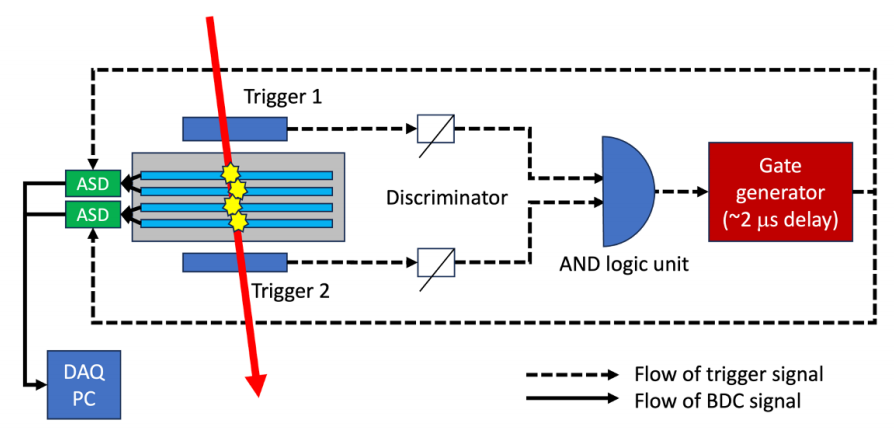}
    \caption{Schematic diagram of the DAQ setup for the test. For the proton beam, upstream and downstream triggers are used, otherwise only upstream trigger 1 is used for the $^{12}$C beam.}
    \label{fig:fig7}
\end{figure}

The Data Acquisition (DAQ) system employed Nuclear Instrumentation Module (NIM) signal modules.
The analog pulse from trigger detectors in each setup was converted to the NIM signal by the discriminator, and the logic unit module made the coincidence of the converted signal from each trigger detector. A gate generator delays the coincidence NIM pulse by 2 $\rm{\mu s}$ before being transmitted to the ASD board as the trigger signal.
Figure~\ref{fig:fig7} provides a schematic view of the DAQ logic. For the $^{12}$C beam only single trigger located in front of the pBDC is used, as referred in previous paragraph. The raw pulses generated by the anode planes were transferred to the processing board via a 32-channel flat cable.
Signal processing was carried out using RPR-010 boards, specifically designed for data acquisition in drift chambers operating at High Energy Accelerator Research Organization (KEK) and produced by HAYASHI-REPIC CO., LTD.
The ASD chip processing optimized the raw signal for time measurement from the pulse. It provides the Time-to-Digital-Converter (TDC) information calculated by the Silicon Transmission Control Protocol (SiTCP) chip based on the charge(Q)-to-Digital-Converter (QDC) values.

The resulting TDC and QDC information was measured within a sampling time of 32 ns.
If the QDC value exceeded the threshold set by the software, TDC time with 1 ns resolution was saved within the period.
The delayed trigger signal entered the ASD board and opened a timing gate window of 320 ns after inversion.
The processed QDC and TDC information was transferred to the DAQ Personnel Computer (DAQ PC) via Unshielded Twisted Pair (UTP) cables.

The tests were conducted using P-10 gas mixture as the operating gas, comprising 90\% argon as the stable component and 10\% methane as the quencher.
The P-10 gas circulated at a pressure of 1 atm and ambient temperature.
The anticipated gain range is expected to be between $10^{4}$ and $10^{5}$, as noted in reference~\cite{ref12}.

\section{Results}
\label{sec:res}

The performance evaluation of the pBDC can be estimated based on the track reconstruction efficiency and position resolution.
The analysis procedure has proceeded consistently for both proton and $^{12}$C beam data, utilizing only information from four $X$ direction layers ($XX'XX'$ configuration) since the data was received only for X direction. The analysis uses the data acquired in High Voltage (HV) of 1150, 1170, and 1190 V for the $^{12}$C beam and 1350, 1360, 1370, 1380, 1390, and 1400 V for the proton beam. 

\subsection{Time distribution and drift velocity}
\label{sec:timedist}

The data analysis begins with measuring the incident particles' drift time of ionized electrons. The drift length of ionized electrons is essential to build particles' tracks. Finally, the combination of drift lengths for 4 $X$ layers enables the construction of the beam trajectories~\cite{ref13}.   
The drift time ($T_{\rm{drift}}$) was extracted by the following equation.

\begin{equation}
T_{\rm{drift}} = T_{\rm{TDC}} - T_{\rm{trig}},
\label{eqn:tTDC}
\end{equation}
where $T_{\rm{TDC}}$ is the time measured at TDC through DAQ system, and $T_{\rm{trig}}$ is the combined time with the delayed time (about 2 $\rm{\mu s}$) set by gate generator in DAQ system and the inversion time set by ASD board.
$T_{\rm{max}}$ is decided as the time of the endpoint of the $T_{\rm{TDC}}$ distribution.

\begin{figure}[htbp]
\centering
\includegraphics[width=.4\textwidth]{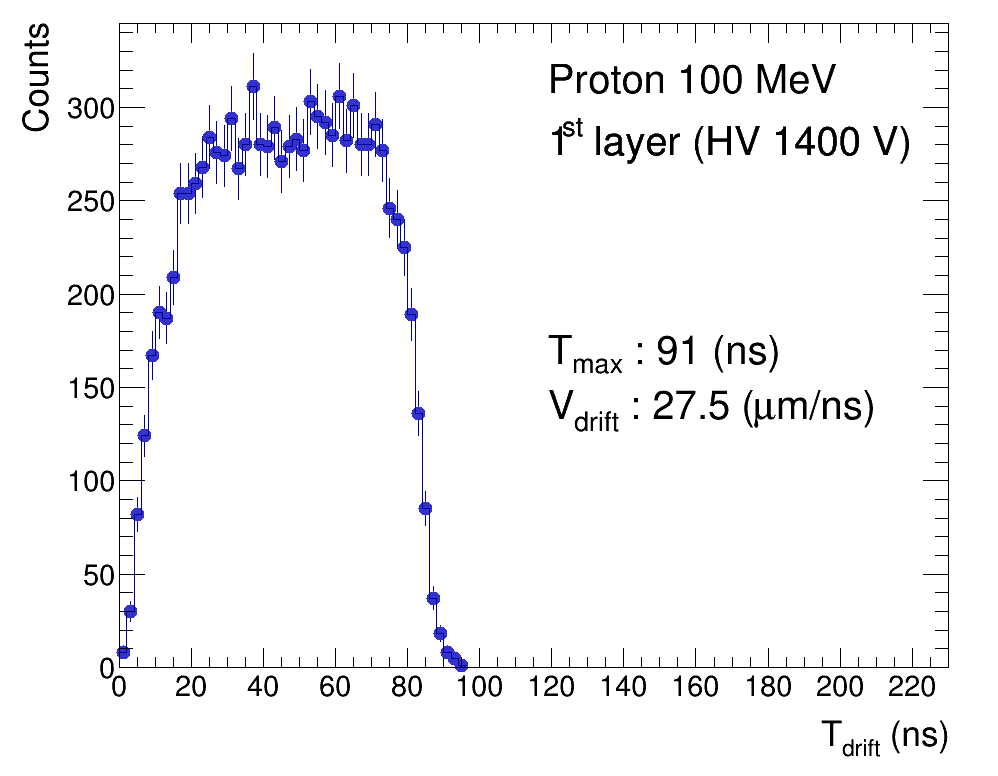}
\includegraphics[width=.4\textwidth]{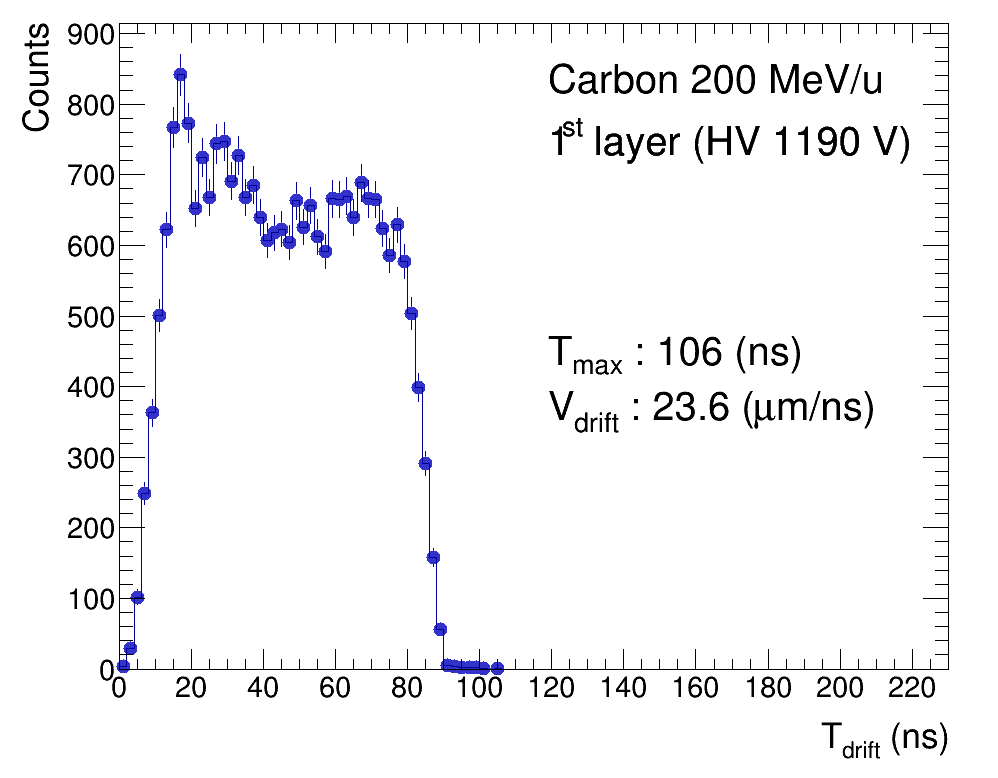}
\caption{\label{fig:fig8} The measured drift time distributions for proton (left) and $^{12}$C (right) beams in the first layer.}
\end{figure}

\begin{figure}[htbp]
\centering
\includegraphics[width=.4\textwidth]{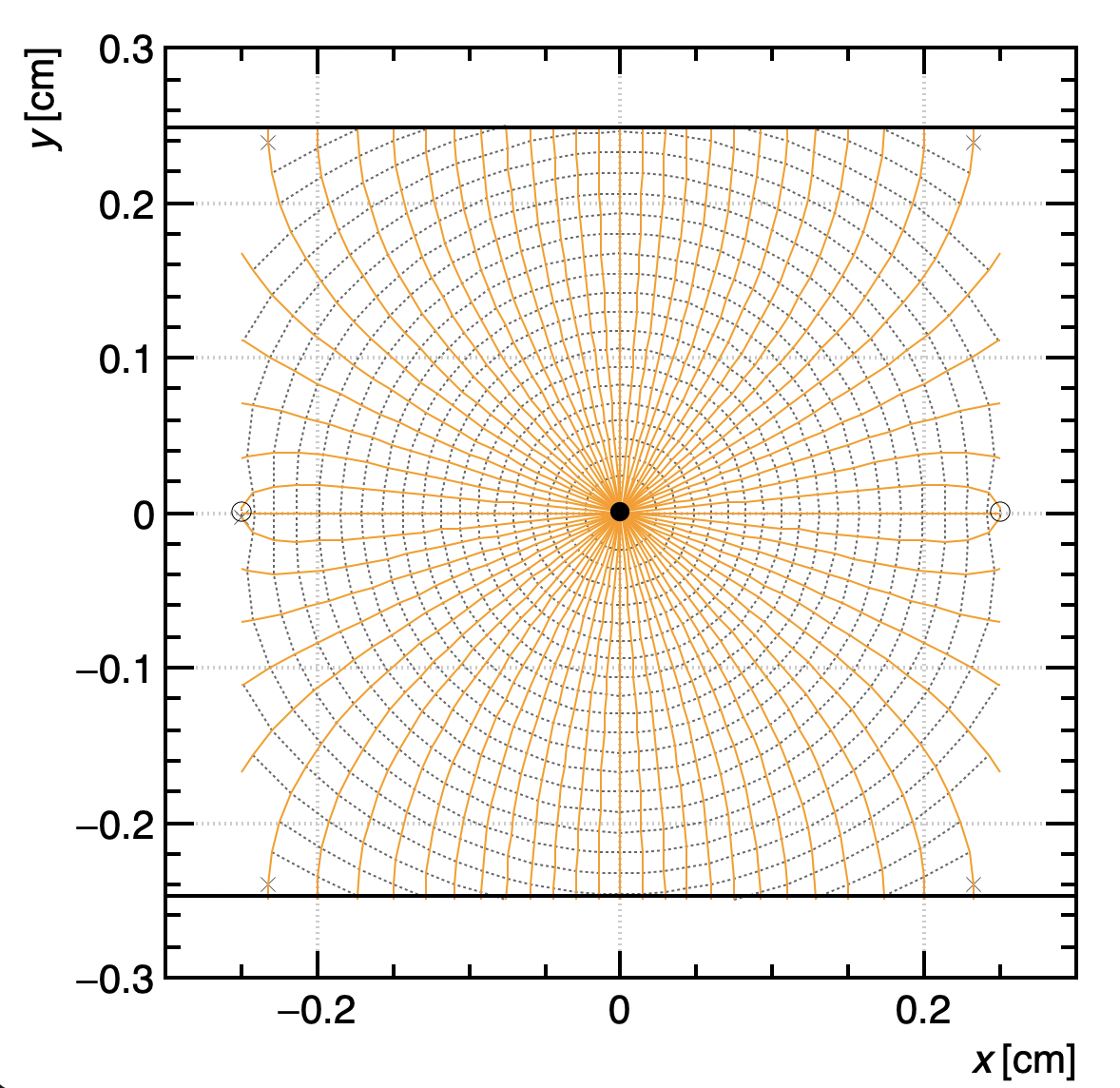}
\includegraphics[width=.4\textwidth]{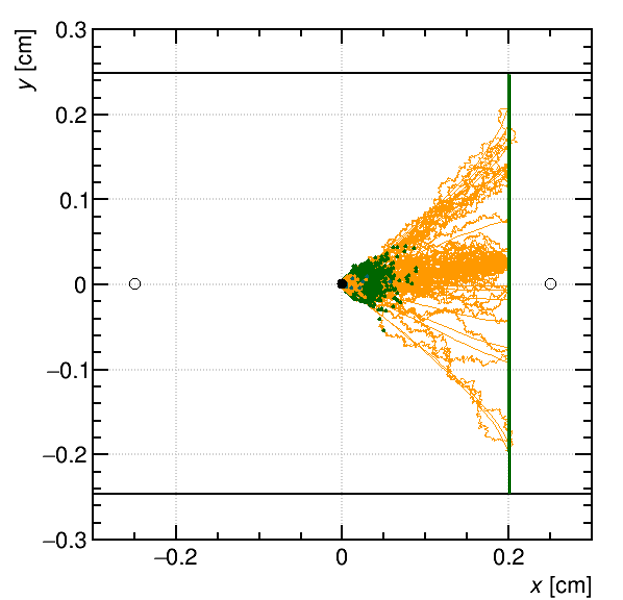}
\caption{\label{fig:fig9} Garfield++ simulation. Electric field map at 1400 V (left) and 100 MeV proton simulation (right). In the right figure, green and yellow lines represent positive ions and electrons, respectively, especially the green vertical line displays the incident proton beam trajectory.}
\end{figure}

\begin{figure}[htbp]
\centering
\includegraphics[width=.4\textwidth]{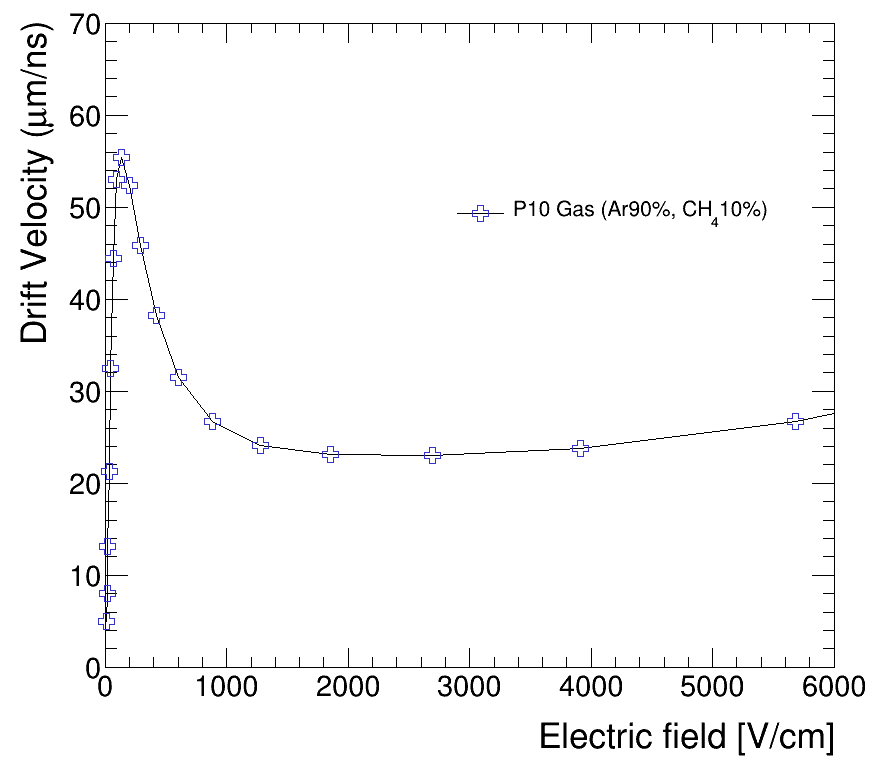}
\includegraphics[width=.4\textwidth]{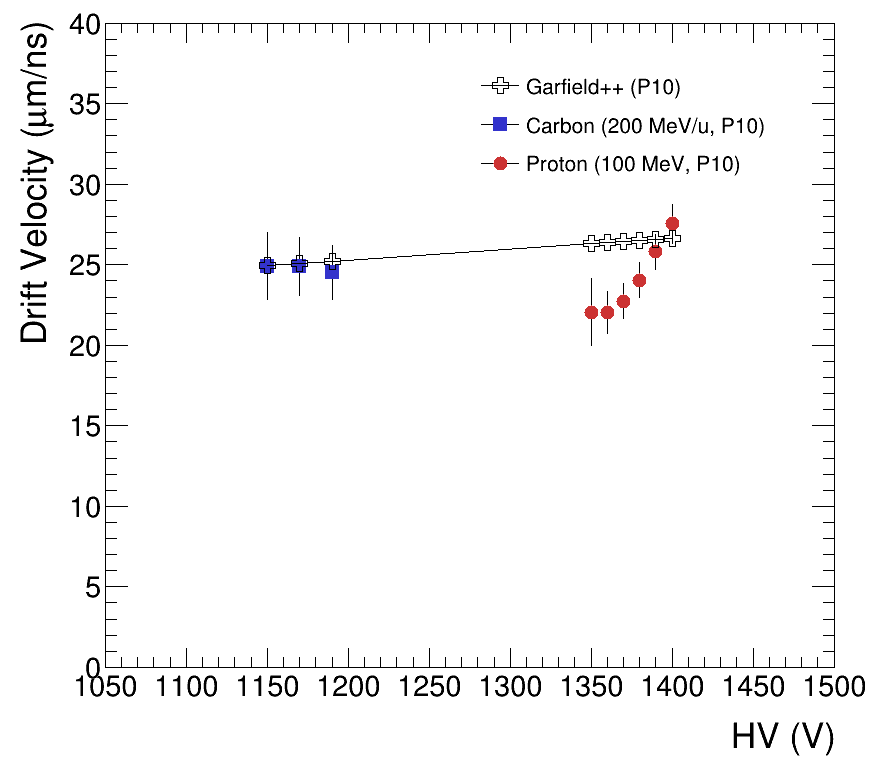}
\caption{\label{fig:fig10} Drift velocity estimated by Garfield++ simulation as the function of applied electric field (left) and comparison between data and Garfield++ simulation (right).}
\end{figure}

The drift velocity ($V_{\rm{drift}}$) is computed individually for each of the four layers, determined by the maximum drift length of 2.5 mm over the maximum drift time ($T_{\rm{max}}$). The overall drift velocity for the pBDC is then obtained by the average of those.
Figure~\ref{fig:fig8} represents the drift time distribution in the first layers for the proton and the $^{12}$C beams. The total error values are computed by systematic uncertainties ($\sim 6 \%$), which are the maximum difference among other layers.
The measured results from the experiments are compared to the simulated one conducted by Garfield++ simulation~\cite{ref14}. Figure~\ref{fig:fig9} shows the electric field map for the case of 1400 V and one example of avalanche simulated situation of 100 MeV proton in the drift chamber. The comparison between data and simulation agreed with each other within uncertainties for the applied voltage region of the $^{12}$C beam, as shown in figure~\ref{fig:fig10}, while for the case of the proton beam, experimental results become closer to simulation results in higher voltage around 1400 V than 1350 V.
These results provide optimum operating voltages depending on the energies and beam species.
The $^{12}$C beam appears to deposit enough energy to measure the expected performance at a lower applied voltage, around 1100 V. However, the proton beam requires increasing the HV up to approximately 1400 V.
At this higher voltage, the experimental results align more closely with the simulated results, indicating that this trend is related to the operating HV range for the detector.
This point will be discussed further in the section on track reconstruction efficiency measurement. 

\subsection{Drift time and length relation}
\label{sec:timelength}

\begin{figure}[htbp]
\centering
\includegraphics[width=.4\textwidth]{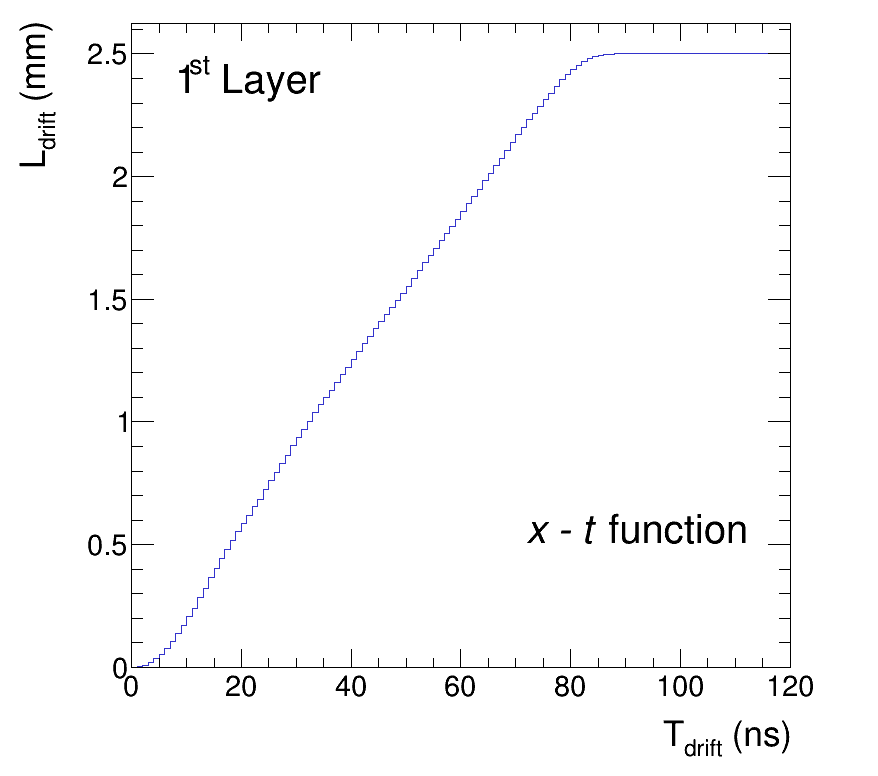}
\caption{\label{fig:fig11} The $x-t$ function derived from the data.}
\end{figure}

The drift time ($T_{\rm{drift}}$) can be converted to the drift length, indicating the distance between the primary ionization point and the fired signal wire's position.
With the assumption of the homogeneous beam distribution, the number of the events within the corrected time interval divided by the number of total events is proportional to the drift length divided by the maximum drift length~\cite{ref15}.
This relation allows the conversion of drift time to drift length.
To achieve a more accurate conversion, the $T_{\rm{drift}}$ distribution of the drift cell around the beam spot center was selected in order to make the beam distributions homogeneous, 
and then the drift time to length conversion function, called $x-t$ function, is obtained. Figure~\ref{fig:fig11} demonstrates the $x-t$ function for the first layer derived from the data.

\subsection{Track reconstruction algorithm}
\label{sec:timelength}

\begin{figure}[htbp]
\centering
\includegraphics[width=1.0\textwidth]{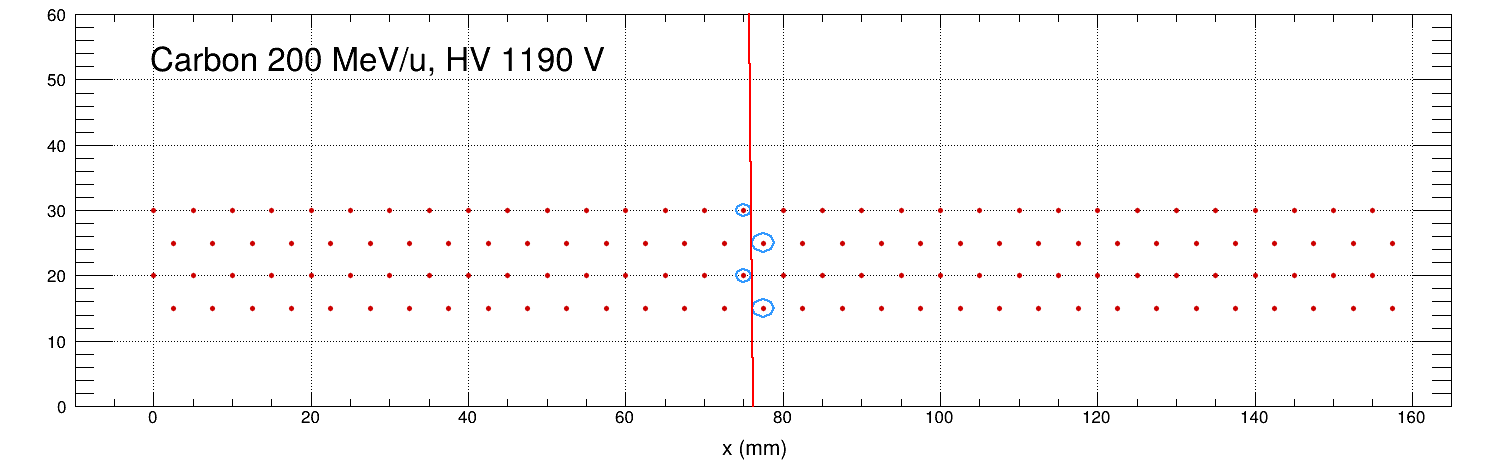}
\caption{\label{fig:fig12} The event display with the reconstructed tracks for the $^{12}$C beam.}
\end{figure}

The $T_{\rm{drift}}$ acquired from each event is converted using the relation from the $x-t$ function for the case of having the corresponding layers and high voltages.
When reconstructing tracks, a circle is drawn with the corresponding drift length converted from drift time by the $x-t$ function as the radius, called a drift circle. The intersected two positions by the drift circle and layers are the candidate points of particle trajectories in the layers. With only one layer, it is hard to determine the exact position due to the intrinsic left-right ambiguity issue in the drift chamber. That is the reason why the drift chamber requires at least four layers to reconstruct the beam track. The two intersected positions (track seed points) in four layers can combine 16 track candidates. After linear fitting, if there are multi track candidate sharing same hit information, only one track is selected by minimizing the chi-square per degree of freedom.
Figure~\ref{fig:fig12} represents the event displays of the example reconstructed tracks using the combinations of the drift lengths in the four layers for $X$ direction for the $^{12}$C beam.

\subsection{Track reconstruction efficiency}
\label{sec:eff}

\begin{figure}[htbp]
\centering
\includegraphics[width=.6\textwidth]{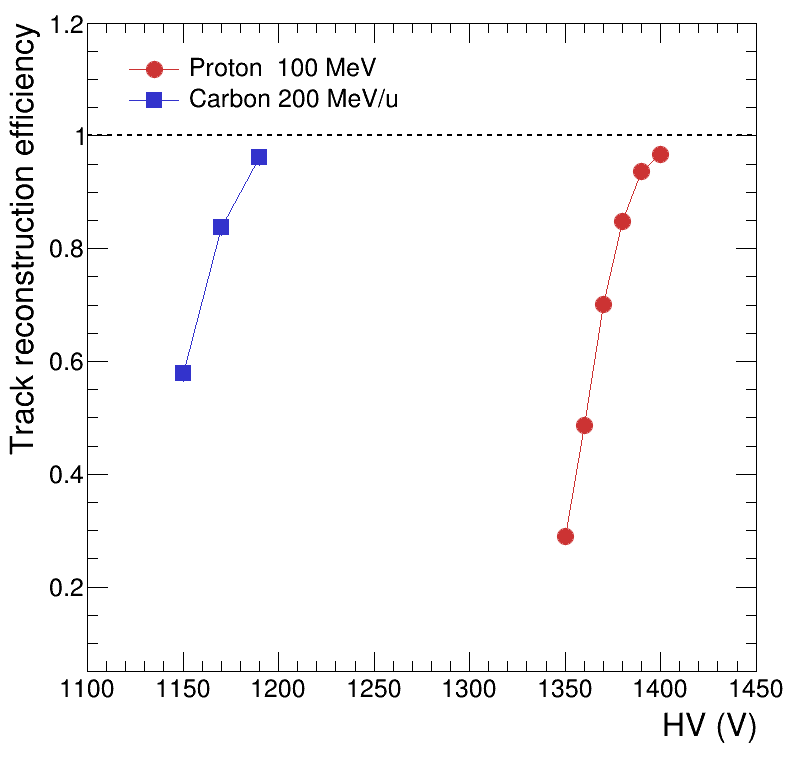}
\caption{\label{fig:fig13} Reconstruction efficiency as the function of applied HV for the $^{12}$C (blue square) and proton (red circle) beams.}
\end{figure}

The tracking efficiency is defined as the ratio of the number of events containing at least one track candidate to the total number of triggered events.
Figure~\ref{fig:fig13} illustrates the reconstruction efficiency as a function of applied HV for $^{12}$C and proton beams.
The reconstruction efficiency reaches a value higher than 95\% at the highest HV applied in this test, suggesting that it is close to the plateau of the working region of the detector, where the simulation and experimental results for the drift velocity measurement agreed within uncertainties as mentioned earlier in the section~\ref{sec:timedist}.

\subsection{Position resolution}
\label{sec:posres}

The position resolution is a crucial parameter for understanding the performance of the tracking detector. 
The position resolution ($\sigma_{\rm{res}}$) is evaluated using the geometric mean of the inclusive and exclusive resolutions~\cite{ref16, ref17}, as expressed in equation~\ref{eqn:res}.  
The inclusive resolution ($\sigma_{\rm{in}}$) is obtained from the average widths of the Gaussian fits to the inclusive residual distributions, which account for the distributions spread across all layers. 
In contrast, the exclusive resolution ($\sigma_{\rm{ex}}$) is estimated from the average Gaussian widths of the exclusive residual distributions after excluding the specific layer under evaluation.

\begin{equation}
\sigma_{\rm{res}} = \sqrt{\sigma_{\rm{in}} \times \sigma_{\rm{ex}}}
\label{eqn:res}
\end{equation}
 
\begin{figure}[htbp]
\centering 
\includegraphics[width=1.0\textwidth]{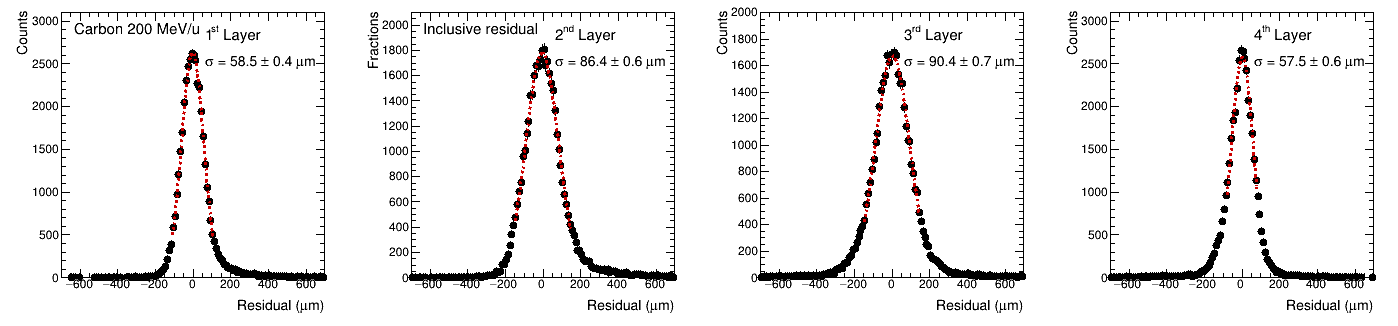}
\includegraphics[width=1.0\textwidth]{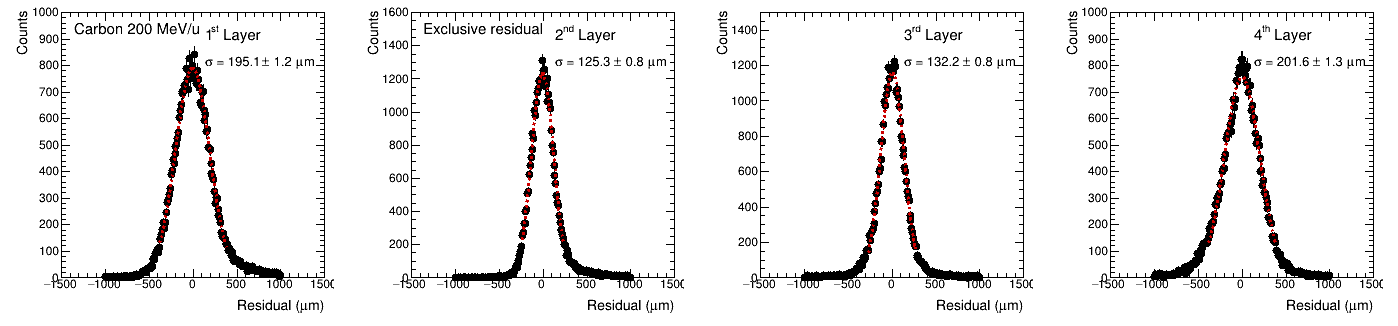}
\caption{\label{fig:fig14} The inclusive (top) and exclusive (bottom) residual distributions for each layer using $^{12}$C beams. The red dotted lines denote the fitted Gaussian functions.}
\end{figure}

\begin{figure}[htbp]
\centering
\includegraphics[width=.4\textwidth]{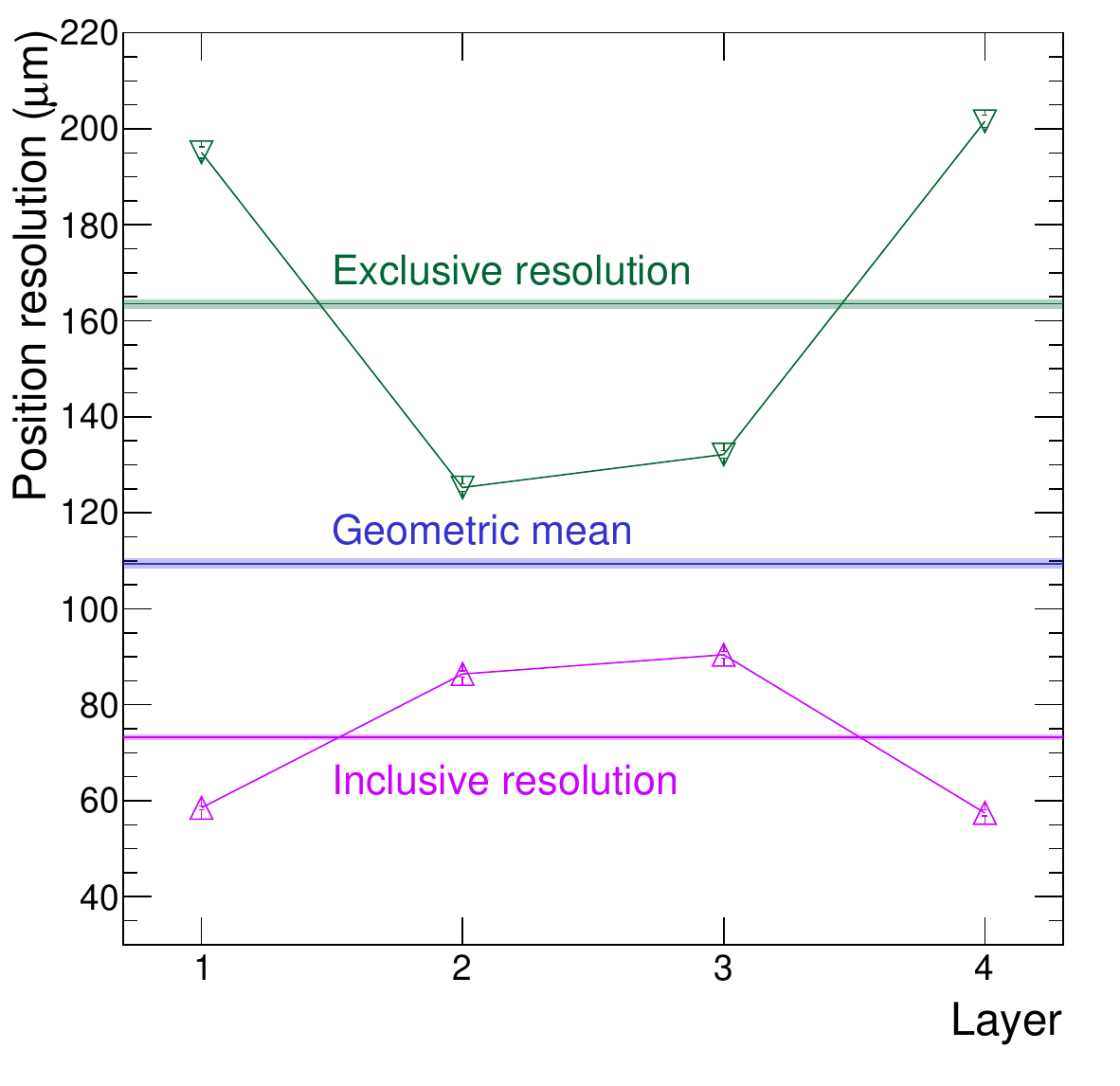}
\includegraphics[width=.4\textwidth]{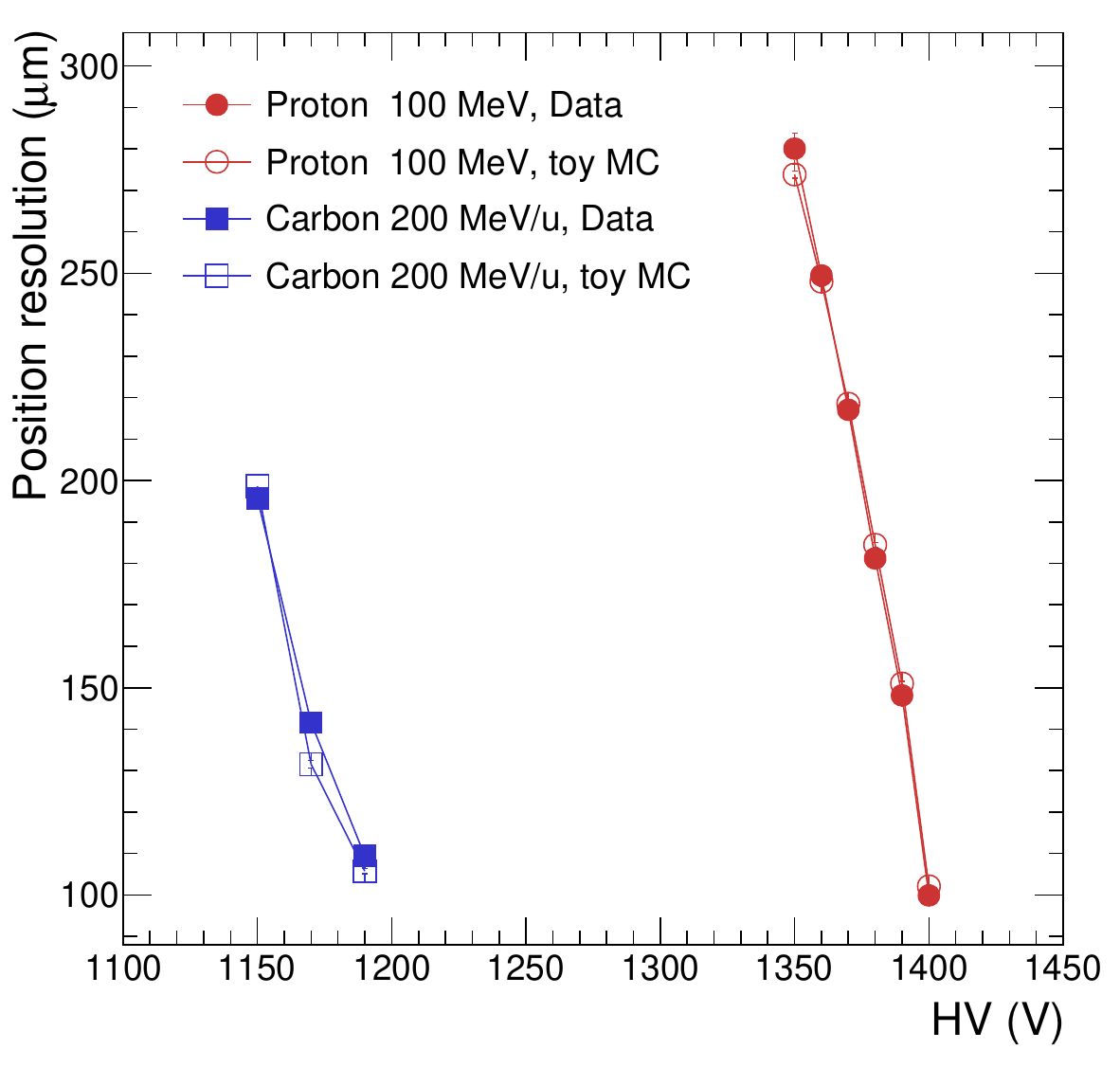}
\caption{\label{fig:fig15} Inclusive and exclusive residuals for each layer are represented by triangle down and up markers, respectively, with the overall resolution indicated by the geometric mean line for $^{12}$C beam at 1190 V (left). Position resolution as a function of applied high voltage for $^{12}$C (blue squares) and proton (red circles) beams. Filled and open markers denote the position resolution evaluated from data and toy MC, respectively. Vertical bars indicate the statistical uncertainties (right).}
\end{figure}

Figure~\ref{fig:fig14} presents the inclusive residual distributions (top panels) and the exclusive residual distributions (bottom panels) for the $^{12}$C beam at 1190 V, evaluated across each layer.
These distributions are used to estimate the inclusive ($\sigma_{\rm{in}}$) and exclusive ($\sigma_{\rm{ex}}$) resolution  evaluations, respectively. 
As shown in the left panel of figure~\ref{fig:fig15}, inclusive and exclusive resolutions are calculated by the average of the residual for each layer, with their geometric mean providing the overall position resolution ($\sigma_{\rm{res}}$).

After setting the detector geometry under the same conditions as performed in the beam test for toy MC simulation used in ref.~\cite{ref17}, 
we consider the case where an example beam track passes through four layers and designate the points where the beam track and the layers intersect as the true track positions. 
These true track positions are used as the mean value, and assuming a Gaussian distribution with a width of the resolution value ($\sigma_{\rm{res}}$) obtained from the pBDC data analysis, randomly single hit was generated for each layer.
Subsequently, four hits obtained from the four layers were randomly combined to create 100,000 toy MC tracks, 
and the analysis was carried out using the same method as used during data analysis. 
The right panel of figure~\ref{fig:fig15} compares the toy MC simulation results with the experimental data, showing good agreement within uncertainties, thereby confirming the validity of the geometric mean method and measured resolutions.

The measured position resolution gets better along with increasing HV as expected.
The best position resolutions are achieved to be 109.4 $\pm$ 1.1 $\mu\rm{m}$ and 99.9 $\pm$ 1.4 $\mu\rm{m}$ for the $^{12}$C beam at 1190 V and proton beam at 1400 V, respectively. 
These results are in close agreement with the required specifications for the LAMPS BDC.

\section{Conclusions}
\label{sec:concl}

The prototype beam drift chamber (pBDC) was developed to assess the performance of the LAMPS BDC, measuring track reconstruction efficiency and position resolution.
It was thoroughly investigated through extensive tests with the $^{12}$C and proton beams provided by the HIMAC facility.
The track reconstruction efficiency and position resolution were evaluated through the reconstructed beam trajectories.
The pBDC exhibited appropriate performance, achieving a track reconstruction efficiency exceeding 95\% and a position resolution below 110 $\rm{\mu m}$ at the expected operational voltages, closely approaching the stringent specifications required.
This study is expected to be the baseline for completing LAMPS BDC.

\acknowledgments

We would like to express our sincere gratitude to the HIMAC institution for their generous support.
This work was supported by the National Research Foundation of Korea (NRF) (2013M7A1A1075764, 2018R1A5A1025563, 2021R1A2C2012584, 2022R1A5A1030700) grants funded by the Ministry of Science and ICT (MSIT).
Additional support was provided by the Global-Learning \& Academic Research Institution for Master’s, PhD Students, and Postdocs (LAMP) Program through the NRF grant funded by the Ministry of Education (Grant No. RS-2024-00442775).


\begin{thebibliography}{99}

\bibitem{ref1}
C. F. von Weisz\"{a}cker, \emph{Zur Theorie der Kernmassen}, \emph{Z. Physik} {\bf 96} (1935) 431.

\bibitem{ref2}
J. M. Lattimer and M. Prakash, \emph{The Physics of Neutron Stars}, \emph{Science} {\bf 304} (2004) 536.

\bibitem{ref3}
Chang Xu1, Bao-An Li, and Lie-Wen Chen, \emph{Relationship between the symmetry energy and the single-nucleon potential in isospin-asymmetric nucleonic matter}, \emph{Eur. Phys. J. A} {\bf 50} (2014) 21.

\bibitem{ref4}
B. Hong et al.,
\emph{Plan for nuclear symmetry energy experiments using the LAMPS system at the RIB facility RAON in Korea}, \emph{Eur. Phys. J. A} {\bf 50} (2014) 49.

\bibitem{ref5}
B. Hong, \emph{Prospects of nuclear physics research using rare isotope beams at RAON in Korea}, \emph{Nucl. Sci. Tech.} {\bf 26} (2015) S20505.

\bibitem{ref6}
B. Hong et al., 
\emph{Development of large acceptance multi-purpose spectrometer in Korea for symmetry energy}, \emph{Nucl. Sci. and Tech.} {\bf 29} (2018) 171.

\bibitem{ref7}
B. Hong et al.,\emph{Status of LAMPS at RAON} \emph{Nucl. Inst. and Meth. in Phys. Res. B} {\bf 541} (2023) 260.

\bibitem{ref8}
J. K. Ahn et al.,
\emph{Overview of the KoRIA Facility for Rare Isotope Beams}, \emph{Few-Body Syst.} {\bf 54} (2013) 197.

\bibitem{ref9}
S. C. Jeong, P. Papakonstantinou, H. Ishiyama, Y. Kim, \emph{A Brief Overview of RAON Physics}, \emph{J. Korean Phys. Soc.} {\bf 73} (2018) 516.

\bibitem{ref10}
H. Shim et al., \emph{Performance of prototype neutron detectors for Large Acceptance Multi-Purpose Spectrometer at RAON}, \emph{Nucl. Inst. and Meth. in Phys. Res. A}, {\bf 927} (2019) 280.

\bibitem{ref11}
A.H. Walenta, \emph{State of the Art of Drift Chambers}, \emph{IEEE Trans.Nucl.Sci.} {\bf 22} (1975) 251-254.

\bibitem{ref12}
Ix-B. G. Ferreira, J. G. Herrera, and L. Villase\~{n}or \emph{The Drift Chambers Handbook, introductory laboratory course}, \emph{J. Phys.: Conf. Ser.} {\bf 18} (2005) 346-361.

\bibitem{ref13}
The ATLAS Collaboration, \emph{Resolution of the ATLAS muon spectrometer monitored drift tubes in LHC Run 2}, \emph{JINST} {\bf 14} (2019) P09011.

\bibitem{ref14}
Garﬁeld++ toolkit, https://garﬁeldpp.web.cern.ch.

\bibitem{ref15}
Dominika Alfs, \emph{Drift Chamber Track Reconstruction for the P349 Antiproton Experiment}, arxiv:1706.09108v1

\bibitem{ref16}
R.K. Carnegie et al.,
\emph{Resolution studies of cosmic-ray tracks in a TPC with GEM readout}, \emph{Nucl. Inst. and Meth. in Phys. Res. A} {\bf 538} (2005) 372.

\bibitem{ref17}
ATLAS collaboration,
\emph{Resolution of the ATLAS muon spectrometer monitored drift tubes in LHC Run 2}, \emph{JINST} {\bf 14} (2019) P09011.



\end{thebibliography}
\end{document}